\documentclass{aa}  

\usepackage{graphicx}
\usepackage[dvipsnames]{xcolor}
\usepackage{svg}
\usepackage{txfonts}
\usepackage{hyperref}
\usepackage{calrsfs}
\usepackage{upgreek}
\usepackage{float}
\DeclareMathAlphabet{\pazocal}{OMS}{zplm}{m}{n}
\newcommand{\greenhyperref}[2]{\hypersetup{linkbordercolor=green}\hyperref[#1]{#2}\hypersetup{linkbordercolor=red}}

\newcommand{\citeyearless}[1]{\citeauthor{#1} \citeyear{#1}}
\usepackage{hyperref}
\begin{document}

   \title{Projection-angle effects when ``observing'' a turbulent magnetized collapsing molecular cloud. {I}. Chemistry and line transfer}

   \author{A. Tritsis
          \inst{1},
          S. Basu\inst{2,3}
          \and
          C. Federrath\inst{4,5}
          }

   \institute{Institute of Physics, Laboratory of Astrophysics, Ecole Polytechnique F\'ed\'erale de Lausanne (EPFL), \\ Observatoire de Sauverny, 1290, Versoix, Switzerland \\
              \email{aris.tritsis@epfl.ch}
         \and
             Department of Physics and Astronomy, University of Western Ontario, London, ON N6A 3K7, Canada
         \and
             Canadian Institute for Theoretical Astrophysics, University of Toronto, 60 St. George, St., Toronto, ON M5S 3H8, Canada
         \and
             Research School of Astronomy and Astrophysics, Australian National University, Canberra, ACT 2611, Australia
        \and
             Australian Research Council Centre of Excellence in All Sky Astrophysics (ASTRO3D), Canberra, ACT 2611, Australia}
   \date{Received date; accepted date}
   \titlerunning{From chemo-dynamics to synthetic observations}
   \authorrunning{Tritsis et al.}
 
  \abstract
   {Most of our knowledge regarding molecular clouds and the early stages of star formation stems from molecular spectral-line observations. However, the various chemical and radiative-transfer effects, in combination with projection effects, can lead to a distorted view of molecular clouds and complicate the interpretation of observations.}
   {Our objective is to simultaneously study all of these effects by creating synthetic spectral-line observations based on chemo-dynamical simulations of a collapsing molecular cloud.}
   {We performed a three-dimensional ideal magnetohydrodynamic simulation of a supercritical turbulent collapsing molecular cloud where the dynamical evolution was coupled to a nonequilibrium gas-grain chemical network consisting of 115 species, the evolution of which was governed by >1600 chemical reactions. We post-processed this simulation with a multilevel nonlocal thermodynamic equilibrium radiative-transfer code to produce synthetic position-position-velocity data cubes of the $\rm{CO}$, $\rm{HCO^+}$, $\rm{HCN}$, and $\rm{N_2H^+}$ ($J = 1\rightarrow0$) transitions under various projection angles with respect to the mean component of the magnetic field. Synthetic polarization maps are presented in a companion paper.}
   {We find that the chemical abundances of various species in our simulated cloud tend to be over-predicted in comparison to observationally derived abundances and attribute this discrepancy to the fact that the cloud collapses rapidly and therefore the various species do not have enough time to deplete onto dust grains. This suggests that our initial conditions may not correspond to the initial conditions of real molecular clouds and cores. We show that the projection angle has a notable effect on the moment maps of the species for which we produced synthetic observations. Specifically, the integrated emission and velocity dispersion of $\rm{CO}$, $\rm{HCO^+}$ and $\rm{HCN}$ are higher when the cloud is observed ``face on'' compared to ``edge on,'' whereas column density maps exhibit an opposite trend. Finally, we show that only $\rm{N_2H^+}$ is an accurate tracer of the column density of the cloud across all projection angles studied.}
   {}


   \maketitle


\section{Introduction}\label{intro}

In order to probe the dynamics of star-forming cores and molecular clouds, we heavily rely on spectral-line observations of various chemical tracers. Such observations are used to probe the turbulent motions within clouds and cores (\citeyearless{1981MNRAS.194..809L}; \citeyearless{2009ApJ...699.1092H}; \citeyearless{2010MNRAS.403.1507B}; \citeyearless{2016ApJ...832..143F}; \citeyearless{2023MNRAS.526..982G}), estimate their angular momenta (e.g., \citeyearless{1993ApJ...406..528G}; \citeyearless{Chen2023}), probe the available gas for star formation (e.g., \citeyearless{2023MNRAS.520.1005J} and references therein), derive the kinetic temperature of the gas, probe the $\rm{H_2}$ number density of these clouds (\citeyearless{2012ApJ...756...12L}; \citeyearless{2023A&A...679A.112T}), and derive information regarding the strength of the magnetic field (\citeyearless{1951Phys.Rev....81...890}; \citeyearless{1953ApJ...118..113C}; \citeyearless{2021A&A...647A.186S}; \citeyearless{2024A&A...684A.212B}). Yet, while observations from such tracers are routinely used, theoretical investigations are still lacking in terms of accurately and simultaneously modeling molecule formation and line-emission processes.

Three-dimensional (3D) simulations that incorporate time-dependent chemistry typically focus on a limited number of species and often involve a restricted set of chemical reactions relevant to $\rm{CO}$ chemistry and/or $\rm{H_2}$ formation (\citeyearless{2010MNRAS.404....2G}; \citeyearless{2012MNRAS.421.2531M}; \citeyearless{2014MNRAS.440..465B}; \citeyearless{2019MNRAS.486.4622C}; \citeyearless{2017MNRAS.472.4797S}; \citeyearless{2020MNRAS.492.1465S}). Such simulations usually focus on scales of a few tens of parsecs where the formation of more complex molecular species than $\rm{CO}$ can generally be ignored. However, such simulations also lack the capacity to provide theoretical predictions regarding the expected observational characteristics of pre-stellar cores where star formation takes place, as high-density tracer molecules are not modeled. Numerical studies that focus on smaller scales and include nonequilibrium chemical modeling (e.g., \citeyearless{2012ApJ...754....6T}; \citeyearless{2015MNRAS.446.3731K}; \citeyearless{2016MNRAS.458..789T}; \citeyearless{2022MNRAS.510.4420T}; \citeyearless{2023MNRAS.521.5087T}) are often limited in terms of dimensionality, and therefore do not capture the full complexity of pre-stellar cores.

An alternative approach to performing chemistry ``on the fly'' is to post-process the dynamical simulation at a specific evolutionary stage using a time-dependent chemical model (\citeyearless{2014MNRAS.437.1662K}; \citeyearless{2023A&A...669A..74G}; \citeyearless{Hu2024}). Such post-processing is less computationally demanding and allows for the exploration of a larger parameter space but these methods do not reflect the significant molecular-abundance variations that can arise from different dynamical evolutionary pathways \citep{2023MNRAS.524.5971P}. To compensate for such shortcomings, other studies have used ``tracer particles'' or a combination of tracer particles and nonequilibrium chemical modeling to predict the molecular abundances of various species (e.g., \citeyearless{2024A&A...685A.112N}, \citeyearless{2024MNRAS.tmp.1424P} and references therein).

Clearly, in all of the aforementioned studies, there is a trade-off between the number of species followed, and/or the methodology through which these species are modeled, and the computational cost. As such, even very detailed studies, where a large number of species are followed using an on-the-fly approach (see for instance \citeyearless{2019ApJ...887..224B}), are forced to exclude more complex molecules. However, given the chaotic nature of chemical networks and the interdependence of species, excluding more complex molecules may result in inaccurate estimates of the abundances of the modeled species. In turn, such inaccuracies can propagate through the radiative-transfer calculations, undermining the overall accuracy of theoretical predictions.

In this study we present the first results from a 3D simulation of a turbulent magnetized collapsing molecular cloud. In addition to modeling the dynamics of the cloud, we employed a nonequilibrium gas-grain chemical network to compute the abundances of 115 chemical species using an on-the-fly approach. Chemical species with up to seven atoms were modeled. We additionally post-processed this simulation with a nonlocal thermodynamic equilibrium (non-LTE) line radiative-transfer code and produced synthetic position-position-velocity (PPV) data cubes. In the present study, emphasis is placed on the qualitative aspects of our simulations and simulated observations, deferring a detailed comparison to actual observations to future research.

Our paper is organized as follows. In Sect.~\ref{numer} we summarize our methods and the numerical tools used to perform the chemo-dynamical simulation and the radiative-transfer calculations. In Sect.~\ref{dynamics} we discuss dynamical and chemical results, and in Sect.~\ref{mockobs} we present synthetic observations of the $\rm{CO}$, $\rm{HCO^+}$, $\rm{HCN}$, and $\rm{N_2H^+}$ ($J = 1\rightarrow0$) transitions under five different projection angles with respect to the mean component of the magnetic field. In a companion paper (\citeyearless{ATSBCF}; hereafter \greenhyperref{paperII}{Paper II}), we present synthetic polarization maps for the same projection angles and analyze their correlation with the simulated PPV cubes presented here. Finally, in Sect.~\ref{discuss}, we summarize our most important results.

\section{Numerical methods and setup}\label{numer}

\subsection{Numerical setup}

To perform our ideal magnetohydrodynamic (MHD) chemo-dynamical simulation, we used the \textsc{FLASH} adaptive mesh refinement code (\citeyearless{2000ApJS..131..273F}; \citeyearless{2008ASPC..385..145D}). We used a hybrid Riemann solver that combines the Roe solver \citep{1981JCP...43..357} for accuracy and the HLLD solver \citep{2005JCoPh.208..315M} for stability, and the unsplit staggered mesh algorithm \citep{1988ApJ...332..659E}. The condition that the divergence of the magnetic field is zero, was consistently met to numerical accuracy at all times. Finally, for the Poisson equation, we used the Multigrid solver \citep{2008ApJS..176..293R}.

We used an ideal gas equation of state, assuming isothermality and setting the temperature equal to $\rm{T} = 10~K$, everywhere in the cloud. The initial number density was set equal to 500 $\rm{cm^{-3}}$, the magnetic field was initially along the $z$ axis and its strength was equal to $7.5~\rm{\mu G}$. As such, the mass-to-flux ratio, in units of the critical value required for the cloud to collapse \citep{1976ApJ...210..326M}, was $\sim$2.3. The total mass of the cloud was equal to $\sim$240 M\textsubscript{\(\odot\)}.

The dimensions of the cloud were equal to 2 pc in each direction. We used a $64^3$ grid with two levels of refinement, such that the resolution of our smallest cell is $\sim8\times10^{-3}$ pc. We used open boundary conditions in all dimensions. We refined our grid using a modified second derivative criterion of the density \citep{1987Comp.Meth.App.Mech.Eng...61..323} and have verified that the Truelove condition \citep{1997ApJ...489L.179T} was consistently met throughout the simulation. We followed the evolution of the cloud up to a ``central'' number density of $10^5~\rm{cm^{-3}}$. Continuing our calculations beyond this point was overly computationally expensive. For reference, more than 2.5 million CPU hours were required to perform this simulation. 

\subsection{Turbulent initial conditions}

We initialized a turbulent velocity field with zero mean momentum, which was generated with the public code \texttt{TurbGen} \citep{2022ascl.soft04001F}, based on the methods described in \cite{2010A&A...512A..81F}. The turbulent velocity field was constructed in Fourier space ($k$-space), with a power-law velocity ($v$) spectrum $dv^2/dk \propto k^{-2}$ in the wavenumber range $2\leq k\leq20$, where $k=2\pi/L$ and $L$ was the size of the computational domain. We generated a natural mixture of modes for the velocity field, with 2/3 of the power in solenoidal modes and 1/3 power in compressible modes, by setting $\zeta=0.5$ in \texttt{TurbGen} (see \mbox{Eqs.~6--9} in \citeyearless{2010A&A...512A..81F}). The standard deviation of the field was set to $\sigma_v = 0.554\,\mathrm{km/s}$, and with the given sound speed of $c_\mathrm{s}=0.18\,\mathrm{km/s}$, this initializes a turbulent velocity field with a sonic Mach number of $\mathcal{M}=\sigma_v/c_\mathrm{s}\sim3$, broadly consistent with the velocity dispersion -- size relation observed in Milky Way clouds (\citeyearless{1981MNRAS.194..809L}; \citeyearless{1992A&A...257..715F}; \citeyearless{2004ApJ...615L..45H}; \citeyearless{2011ApJ...740..120R}). This choice of initial turbulent velocity field approximates the typical turbulent conditions observed in pre-stellar clouds and cores.

\subsection{Chemical modeling}

The inclusion of a chemical network in dynamical simulations is imperative if we wish to have realistic chemical conditions. Having realistic chemical conditions is in turn crucial for radiative-transfer post-processing and ultimately for performing a one-to-one comparison to real observations. Here, we used the same chemical network as in \cite{2022MNRAS.510.4420T} and \cite{2023MNRAS.521.5087T} (see also \citeyearless{2016MNRAS.458..789T}) consisting of 115 species, 48 out of which were ions and 37 were ice-mantle species. At each grid point and for every timestep, the evolution of these species was governed by $\sim$1650 chemical reactions. These reactions encompassed a range of processes, including freeze-out, thermal desorption, photodissociation, photoionization, cosmic-ray ionization, cosmic-ray induced photo processes, and grain-surface chemical reactions. Reaction rates were taken from the \textsc{UMIST} database \citep{2013A&A...550A..36M}. In a future work, we plan to update our chemical network to incorporate the latest reaction rates provided by \cite{2024A&A...682A.109M}. The initial elemental abundances can be found in Table 1 of \cite{2022MNRAS.510.4420T}.

The visual extinction that enters in the determination of the reaction rates of photo-related processes was calculated using an on-the-fly approach based on the six-ray approximation. Following \cite{2012MNRAS.421..116G} the ``3D'' $A_v^{3D}$ is given by
\begin{equation}\label{visualotf}
A_v^{3D} = -\frac{1}{f} log\Bigg[\frac{1}{6}\sum_1^6 e^{-fA_{v, i}}\Bigg] + C,
\end{equation}
where $A_{v, i}$ is the visual extinction along each direction in each of the principal axes of our numerical grid. The inclusion of the factor $f$ was based on the scaling of the photodissociation rate with visual extinction, as $e^{-fA_v}$. Here, we adopted $f = 2.3$ which represents an average scaling factor of all species in the chemical network. To convert between column density $N_{\rm{H_2},i}$ and visual extinction $A_{v, i}$ we followed \cite{2010ApJ...721..686P} and used $N_{\rm{H_2}}/A_{v, i} = 9.4\times10^{20}$ $\rm{cm^{-2}}~\rm{mag^{-1}}$. Given that the simulation presented is intended to represent a subregion of a molecular cloud, a constant offset $C$ was also used in Eq.~\ref{visualotf}. The addition of an offset in Eq~\ref{visualotf} also ensures that there are no issues at the boundaries of our simulation, where the computed value of $A_v^{3D}$ would otherwise be $\ll1$. Here, we adopted a value of $C = 1.5~\rm{mag}$ (see for instance Table 1 from \citeyearless{2011A&A...530A..64K} for an estimate of the average visual extinction in nearby molecular clouds).

A similar expression as Eq.~\ref{visualotf} was also used for the calculation of the cosmic-ray ionization rate. To convert between the column density and the cosmic-ray ionization rate we used \citep{1983PThPh..69..480U}
\begin{equation}\label{zetaotf}
\zeta_{cr} = \zeta_0~\rm{exp}(-\Sigma_{\rm{H_2}}/96~g~cm^{-2}),
\end{equation}
where $\zeta_0 = 1.3\times10^{-17}~\rm{s^{-1}}$ is the ``standard'' cosmic-ray ionization rate \citep{1998ApJ...499..234C}, and $\Sigma_{\rm{H_2}}$ is the mass column density. However, given the range of column densities that we followed here ($< 10^{23} \rm{cm^{-2}}$; see Fig. 1 in \greenhyperref{paperII}{Paper II}), the exponential factor on the right-hand side of Eq.~\ref{zetaotf} was always very close to unity, and only very small deviations ($\ll 1\%$) from the standard value were found.

\subsection{Line radiative-transfer simulations}\label{RTSetup}

To create synthetic spectral-line observations we used the non-LTE multilevel radiative-transfer code $\textsc{PyRaTE}$ (\citeyearless{2018MNRAS.478.2056T}; \citeyearless{2024A&A...692A..75T}). $\textsc{PyRaTE}$ uses a ray-tracing approach to solve the radiative-transfer problem with the level populations calculated using a $\Lambda$-iteration. While doing so, we took into account that photons emitted from one region of the cloud would interact with another region, provided that the relative velocity between the two regions is less than the thermal linewidth. Consequently, this approach allowed us to account for the coupling of the level populations over the entire cloud (see also the discussion in Appendix C in \citeyearless{2024A&A...692A..75T}). To save computational time, however, we neglected the hyperfine structure of molecules, where relevant.

All of our radiative-transfer calculations were performed under non-LTE conditions, assuming five rotational levels and we post-processed our simulation at a resolution of $128^3$ grid cells. The contribution from the Cosmic Microwave Background radiation was taken into account when computing the level populations. Collisional and Einstein coefficients were taken from the $\textsc{LAMBDA}$ database \citep{2005A&A...432..369S}. For all of our radiative-transfer calculations we assumed a spectral resolution of 0.05 $\rm{km ~s^{-1}}$ and 64 points in frequency space. Finally, intensities were converted to brightness-temperature units and noise was added to the spectra such that the signal-to-noise ratio in the center of the cloud was equal to 20 in all modeled PPV data cubes. 



\begin{figure}
\includegraphics[width=1.\columnwidth, clip]{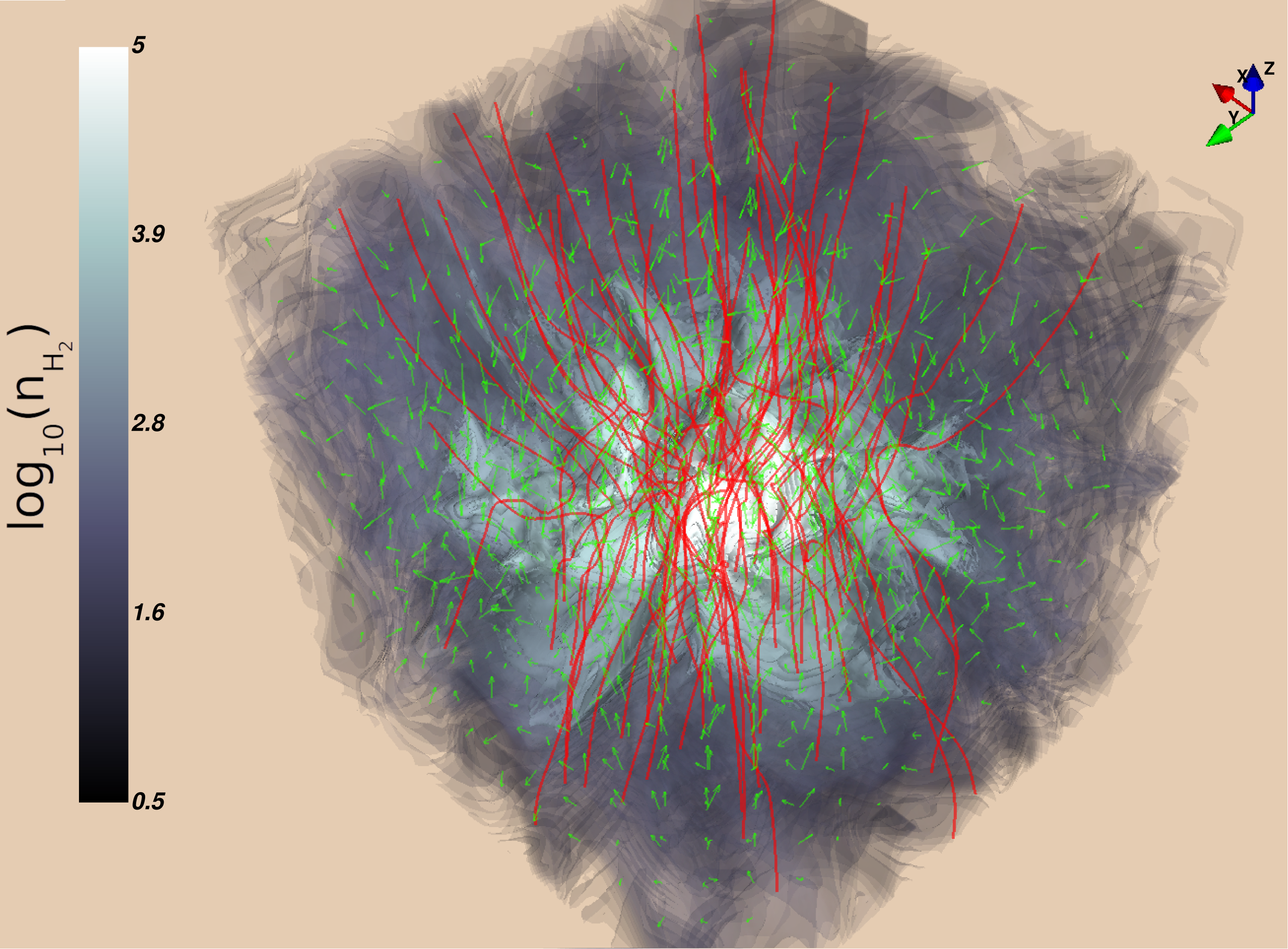}
\caption{Isometric 3D view of the $\rm{H_2}$ number-density distribution (color-coded isosurfaces) within the simulated cloud. Red lines show the magnetic field lines and green vectors show the velocity field in the cloud.
\label{dens3D}}
\end{figure}

\begin{figure*}
\includegraphics[width=2.\columnwidth, clip]{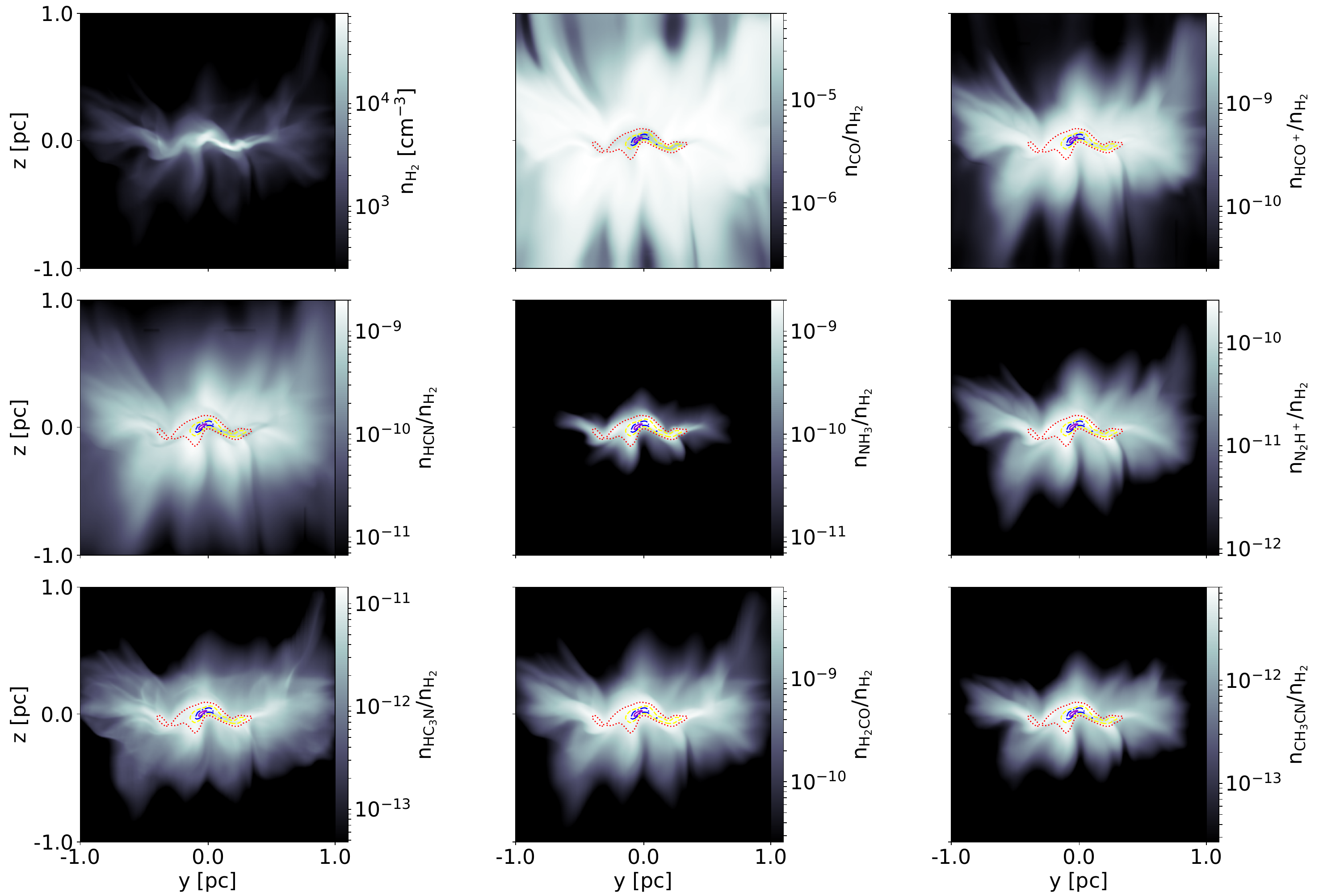}
\caption{Averaged slices (within $0.05$ pc) along the $x$ axis, centered at the location of the maximum number density in the cloud, showing the spatial distribution of various molecules. In the upper left panel we show the $\rm{H_2}$ number density, and in the rest of the panels we show the abundances of various, commonly observed species. The dotted red, dash yellow, dash-dotted blue, and solid violet contours represent 30, 50, 70, and 90\% of the maximum 3D visual extinction within the cloud, respectively.
\label{SpatialChem}}
\end{figure*}

\section{Results}\label{Results}

Throughout this manuscript, we use $x, y, z$, to denote our axes when we show results associated with 3D quantities (e.g., number density), and $\xi$ and $\eta$ for results associated with synthetic observations (i.e., moment maps from our synthetic PPV cubes), including when the line of sight (LOS) coincides with one of the principal axes of the grid. In all images associated with synthetic observations, we annotate the azimuthal and polar angles ($\theta$ and $\phi$, respectively) under which the cloud is ``observed'' such that there is no confusion regarding the projection angle. A schematic illustration of the angles and definitions associated with synthetic observations can be found in Fig. 1 of \greenhyperref{paperII}{Paper II}.

\begin{figure*}
\includegraphics[width=2.\columnwidth, clip]{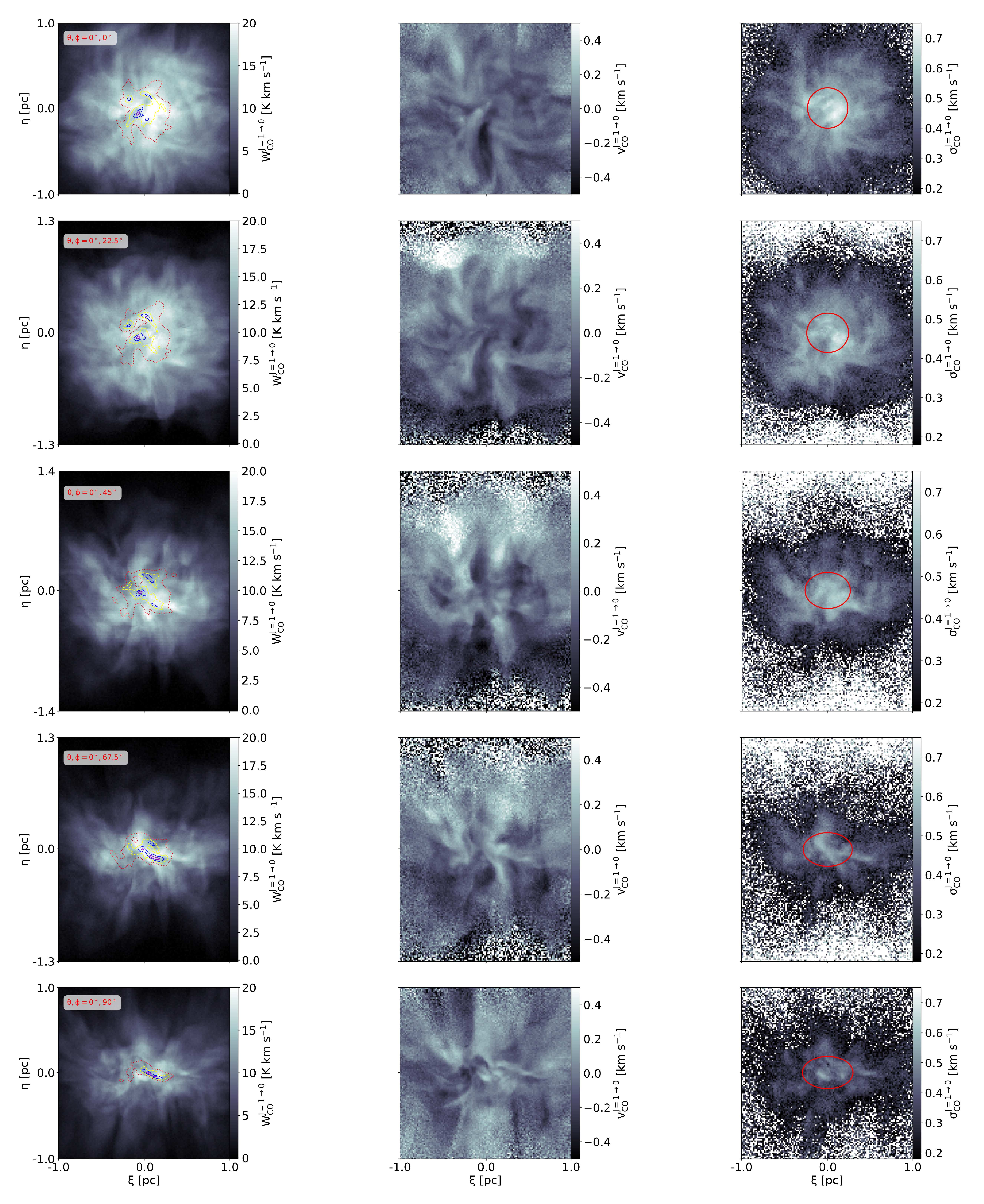}
\caption{Moment maps from the $\rm{CO} ~(J = 1\rightarrow0)$ transition from our chemo-dynamical simulation under different projection angles. In the left, middle and left columns we show the zeroth, first and second moment maps, respectively. In the upper row the mean component of the magnetic field is along the LOS, whereas in the bottom row the mean magnetic field lies in the plane of the sky. Intermediate angles are shown in the second, third and fourth rows (the azimuthal and polar angles are indicated in the upper left corners of the zeroth moment maps). The red dotted, yellow dash, blue dash-dotted, and solid violet contours represent 30, 50, 70, and 90\% of the maximum column density, respectively. The red ellipses in the right column show the region used to extract average values as a function of the projection angle (see text and Fig.~\ref{PrjDep}).
\label{COMM}}
\end{figure*}

\subsection{Chemo-dynamical results}\label{dynamics}

In Fig.~\ref{dens3D} we present a 3D isometric view of our simulated cloud when the central number density is $\sim10^5~\rm{cm^{-3}}$. The magnetic field is depicted using red lines and the green vectors show the velocity field. The cloud exhibits a complex structure, but despite the fact that it is supercritical and super-Alfv\'enic, it has still contracted along the magnetic-field lines (i.e., along the $z$ axis). Consequently, a flattened structure with ``mean density'' of $\sim10^4~\rm{cm^{-3}}$ has formed with its short axis oriented parallel to the magnetic field. That is, the structure itself lies approximately in the $x$-$y$ plane. For reference, the orientation of the coordinate axes is indicated in the top right corner of Fig.~\ref{dens3D}.

\subsubsection{Qualitative features in molecular abundances}

In Fig.~\ref{SpatialChem} we present molecular-abundance slices along the $x$ axis. The slices are averaged within $\rm 0.05$ pc and centered at the location of the maximum $\rm{H_2}$ number density. The calculated molecular abundances exhibit certain general qualitative trends, observed in molecular clouds and pre-stellar cores. The $\rm{CO}$ abundance (upper row, middle panel) decreases by almost an order of magnitude from the outskirts of the cloud to the location of maximum number density where it is depleted onto dust grains. $\rm{HCO^+}$ (upper row, right panel) is significantly more concentrated toward the densest regions of the cloud but it is still abundant at $\rm{H_2}$ number densities of $10^3~\rm{cm^{-3}}$. Similarly, $\rm{HCN}$ (middle row, left panel) also appears to be significantly abundant even at the lower-density regions of the cloud of $\sim 10^3~\rm{cm^{-3}}$ (e.g., \citeyearless{2017A&A...599A..98P}; \citeyearless{2017A&A...605L...5K}; \citeyearless{2021A&A...646A..97T}; \citeyearless{2023MNRAS.520.1005J}). In terms of the ``traditional'' high-density tracers, $\rm{N_2H^+}$ is visibly more extended than $\rm{NH_3}$ (middle row, right and middle panels, respectively), as observations of dense cores also suggest (e.g., \citeyearless{2002ApJ...569..815T}; \citeyearless{2014ApJ...790..129C}). We note here that such differences in the abundance distributions between these two species could potentially account for the fact that the $\rm{N_2H^+}$ spectra have been found to have larger linewidths than those of $\rm{NH_3}$ \citep{2021ApJ...912....7P}. The abundance distributions of $\rm{HC_3N}$ and $\rm{H_2CO}$ (bottom row, left and middle panels, respectively) also seem to be in line with observations in terms of the fact that they peak off center in comparison to the $\rm{H_2}$ number density \citep{2006A&A...455..577T}. 

An important trend that can be observed throughout Fig.~\ref{SpatialChem} is that the abundance distributions of various species appear to be significantly affected by (and often follow) the 3D visual extinction (colored contours in Fig.~\ref{SpatialChem}) which peaks off center in comparison to the $\rm{H_2}$ number density. In 3D space, as well as in the slices presented in Fig.~\ref{SpatialChem}, the distance between the density and visual extinction peaks is $\sim$0.22 pc. Apart from the effect of depletion, this trend may also be a potential reason why certain species peak off center with respect to the continuum peak.

Another notable feature in Fig.~\ref{SpatialChem} is that the small ``filaments'' evident in the $\rm{H_2}$ density distribution (upper left panel) are neither evident in the visual extinction (colored contours) nor in the species. The differences in the filamentary structures seen in the $\rm{H_2}$ density and extinction map arise from the averaging process inherent in the six-ray approximation used to calculate $A_v^{3D}$ (Eq.~\ref{visualotf}). While the $\rm{H_2}$ density reflects sharp variations caused by the turbulent flow, $A_v^{3D}$ is influenced by contributions from all six directions, which effectively ``smooths out'' these sharp changes. Therefore, this smoothening of the small filamentary structures occurs because $A_v^{3D}$ is not localized but rather determined by the surrounding structure from multiple directions.

This smoothing effect of $A_v^{3D}$ also indirectly propagates to the molecular-abundance distributions. However, the fact that other molecular species exhibit different or smoother abundance distributions, lacking the small-scale filaments observed in the $\rm{H_2}$ number density, is not solely due to photo-related processes. These differences arise from the complex interdependence between species in the chemical network, combined with varying timescales and the chemical history of the cloud. In the cloud center, for the time shown in Fig.~\ref{SpatialChem}, and for the chemical history of this specific simulated cloud, $\rm{HCN}$ is primarily produced through reactions involving $\rm{HCNH^+}$ and electrons, and destroyed by reactions with $\rm{H_2CO^+}$. $\rm{CO}$ is produced via reactions involving $\rm{HCO^+}$ and electrons, and destroyed by freezing onto dust grains or reacting with $\rm{H_3^+}$. Only $\rm{N_2H^+}$ is more closely tied to the $\rm{H_2}$ density, since $\rm{H_2}$ is directly involved in its production, although the dominant production mechanism still involves reactions between $\rm{H_3^+}$ and $\rm{N_2}$. For the time shown in Fig.~\ref{SpatialChem}, all of these reactions occur on much shorter timescales than the dynamical timescale of the cloud, meaning that chemistry can, in principle, respond rapidly to the formation of the small filaments observed in the $\rm{H_2}$ density distribution. However, this is not the case as the formation and destruction of these species is not typically directly tied to the $\rm{H_2}$ density, and even when it is (as in the production of $\rm{N_2H^+}$), the other reactant (e.g., $\rm{N_2^+}$ in the case of $\rm{N_2H^+}$) must also reflect to some extent the same features.

\subsubsection{Comparison of modeled and observed abundances}

From a quantitative standpoint, the picture is far less clear. Specifically, the abundances of some species are over-predicted compared to what observations suggest. \cite{2002ApJ...570L.101B} and \cite{2020A&A...635A.188L} both estimated the $\rm{N_2H^+}$ abundance to be $\sim3\times10^{-11}$ in the pre-stellar cores L1512 and B68, respectively. Such values are $\sim$one order of magnitude lower than what is calculated in our simulation. A similar argument can be made for the abundances of $\rm{HCO^+}$ (see for instance the abundance estimates in Table 9 of \citeyearless{2017A&A...599A..98P}). \cite{2006A&A...455..577T} estimated the fractional abundance of this molecule to be $\sim3\times10^{-9}$ and $\sim1.5\times10^{-9}$ in the pre-stellar cores L1498 and L1517B, respectively. Such values are again lower by a factor of $\sim$5 than those calculated in our model. For the same cores, the abundance of $\rm{H_2CO}$ is also estimated to be $\sim$one order of magnitude lower than the predictions of the model, while the abundance of $\rm{HC_3N}$ is in agreement with observational estimates (see also the discussion below). In line with this trend, the observational estimates for the abundance of $\rm{CH_3CN}$ in the pre-stellar cores L1544, L1498 and L1517B are also lower by a factor of $\sim$5-10 compared to model predictions \citep{2023MNRAS.519.1601M}.

Some of these differences may arise due to the specific details of the chemical network employed. Variations in reaction rates across chemical databases, and differences in the number of reactions or species considered, can result in abundance variations of up to a factor of $\sim$5 for certain molecules (e.g., \citeyearless{2012ApJS..199...21W}, Fig. 1). However, in our case, we observe differences often spanning orders of magnitude between the abundances of several molecules in our simulated cloud and those observed. This indicates that the issue likely lies not in the calculation of abundances but rather in the initial conditions of the simulated cloud, which may not accurately represent real clouds. A slower cloud collapse, resulting from different initial conditions, could potentially reduce these differences. For instance, an initially stronger magnetic field such that the cloud was subcritical, and/or a higher initial sonic Mach number could provide additional support against gravity. This would in turn allow chemistry more time to evolve at lower densities and visual extinctions.

The fundamental role of initial conditions in shaping the chemical evolution of molecular clouds becomes evident if we compare the abundances of $\rm{CO}$ and $\rm{HCO^+}$ in our modeled cloud with those obtained from subcritical simulations, where collapse occurs more slowly. For instance, in their 2D subcritical models, \cite{2022MNRAS.510.4420T} found $\rm{CO}$ and $\rm{HCO^+}$ abundances that were one to two orders of magnitude lower than those calculated here for the same central density (see their Figure 10). This is to be expected, as the cloud presented here reaches this evolutionary stage very rapidly (i.e., within $\sim1.8$ Myrs) and, as a result, these species do not have enough time to deplete onto dust grains. Specifically, the depletion timescale of $\rm{CO}$ in our chemical network at a density of $10^4~\rm{cm^{-3}}$ is 0.56 Myrs (see also \citeyearless{2007ARA&A..45..339B}). Comparing the depletion timescale with the free-fall time at the same density, we find that $\tau_{depl}/t_{ff} = 1.7$. In contrast, initially subcritical modeled clouds can spend a significant portion of their lifetime (4-8 Myrs) at densities of $10^3 - 10^4~\rm{cm^{-3}}$ (see Figure 1 in \citeyearless{2023MNRAS.521.5087T}). At these densities, the depletion timescale is shorter by a factor of 2-10 compared to the dynamical timescale, which, in this case, is the ambipolar-diffusion timescale. A similar trend, where the abundances of various species were higher in supercritical models compared to subcritical ones was also observed in the chemo-dynamical simulations by \cite{2012ApJ...754....6T}. 

While in most instances the abundances from the model are larger than observational estimates, there are some cores and clouds where the calculated abundances are in fair agreement with observations or even lower. For example, the $\rm{HC_3N}$ abundance calculated in Lupus starless cores is in very good agreement with model predictions \citep{2019MNRAS.488..495W}. In relative terms, the abundance of $\rm{HC_3N}$ over that of $\rm{NH_3}$ also appears to be in rough agreement with observations \citep{2021SCPMA..6479511X}, albeit the uncertainties span $\sim$two orders of magnitude and therefore no robust conclusions can be drawn. The $\rm{N_2H^+}$ abundance in the pre-stellar cores L1498 and L1517B \citep{2006A&A...455..577T} is also in very good agreement with the predictions of the model, while for the same cores, the estimated $\rm{NH_3}$ abundance is larger than what is calculated from our nonequilibrium chemical model.

From the above discussion it becomes clear that there are order-of-magnitude uncertainties when performing a comparison between model predictions of molecular abundances and observational estimates. Additionally, the results from our model might agree with observations of some cores and clouds and disagree with others. As a result, no definitive conclusion can be drawn using this single model, albeit the present comparison suggests that the initial conditions of our simulated cloud may not accurately represent real clouds. Therefore, a careful dedicated study is required with an extensive literature review of observational abundance estimates of many molecular species over a range of $\rm{H_2}$ densities. A statistical comparison with the results from many different chemo-dynamical models with different initial and physical conditions can then reveal invaluable information regarding the histories of these clouds. In the future, we will revisit this issue with a broader suite of simulations, including subcritical models under nonideal MHD conditions.

\subsection{Synthetic line-emission observations}\label{mockobs}

\begin{figure}
\includegraphics[width=1.\columnwidth, clip]{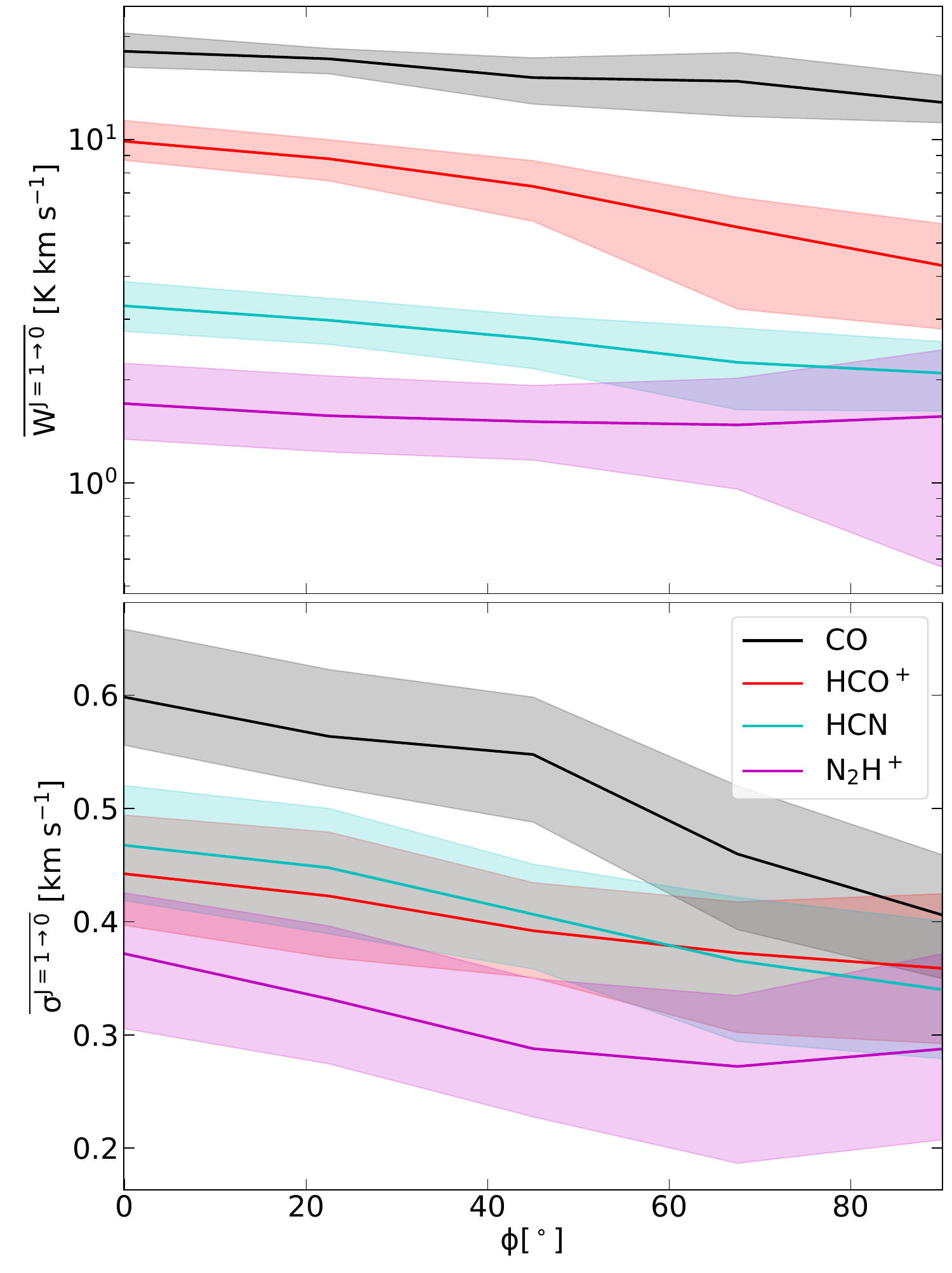}
\caption{Dependence of the average of the zeroth and second moment maps (upper and lower panels, respectively) on the projection angle for each of the species for which we have produced synthetic observations. Averages are computed within the red ellipses shown in the right column of Fig.~\ref{COMM}. Shaded regions show the 16th and 84th percentiles of the distributions of values within the ellipses shown in the right column of Fig.~\ref{COMM}. For all species, we observe a decrease in the derived velocity dispersion with increasing polar angle. The same trend is true for the velocity-integrated emission, apart from $\rm{N_2H^+}$, for which the average of the zeroth-moment maps remains roughly constant.
\label{PrjDep}}
\end{figure}

In Fig.~\ref{COMM} we show moment maps from simulated PPV cubes of the $\rm{CO}$ $J = 1\rightarrow0$ transition. Moment maps from simulated PPV cubes of the $\rm{HCO^+}$, $\rm{HCN}$, and $\rm{N_2H^+}$ $J = 1\rightarrow0$ transitions are shown in Figs.~\ref{HCOpMM},~\ref{HCNMM}, and~\ref{N2HpMM}, respectively. In the left column we show the velocity-integrated emission maps, in the middle column we show the first-moment maps, and finally in the right column we show the second moment maps. Different rows correspond to simulated PPV cubes under different projection angles starting from ``face on'' ($\phi = 0^\circ$; upper row) to ``edge on'' ($\phi = 90^\circ$; bottom row). The colored contours overplotted on top of the zeroth moment maps for each projection angle correspond to the column density with contour levels set equal to 30, 50, 70, and 90 \% of the maximum column density in each specific projection angle (dotted red, dash yellow, dash-dotted blue and solid violet lines, respectively). Column density maps for the same projection angles are presented in \greenhyperref{paperII}{Paper II}. 

The red ellipses in the right column of Fig.~\ref{COMM} show different regions selected to extract mean values of the zeroth and second moment maps as a function of the projection angle. As the polar angle increases from 0$^\circ$ to 90$^\circ$, the shape of these regions transitions from a circle to an ellipse with progressively greater eccentricity to account for the observed morphology of the cloud. Importantly, we ensured that the area of these regions remains constant across all projection angles. The size of these regions was determined based on two criteria. First, a physical criterion, as our primary interest lies in the inner parts of the cloud, where the emission from the rest of the simulated species is also concentrated (i.e., see Fig.~\ref{N2HpMM}). Secondly, a statistical criterion, as the zeroth moment map values within these regions exhibit a roughly Gaussian distribution (for most projection angles and molecules). We verified that changing the size and/or the shape of these regions does not qualitatively impact our results. Our results for the mean values of the zeroth and second moment maps are shown in the upper and lower Fig.~\ref{PrjDep}, respectively. With the black, red, cyan and magenta lines we show our results for CO, $\rm{HCO^+}$, HCN, and $\rm{N_2H^+}$, respectively. The shaded regions in each panel show the 16th and 84th percentiles of the respective distributions inside the red regions in the left column of Fig.~\ref{COMM}.

As evident from Fig.~\ref{COMM}, the $\rm{CO}$ emission does not probe the densest parts of the cloud and systematically peaks off center in comparison to the column-density peaks. While this statement holds true for all projection angles, the effect is more prominent for polar angles close to $90^\circ$ where two quasi-parallel (to each other and to the column-density filament) elongated structures are observed. Additionally, the average integrated emission systematically decreases with an increasing value of the polar angle by approximately $\sim$30 - 50\% from face on to edge on. This trend is more clearly evident in the upper panel of Fig.~\ref{PrjDep} (black line and shaded region). Similarly, this trend in mean integrated intensity is observed in the zeroth moment maps of $\rm{HCN}$ and $\rm{HCO^+}$ (see Figs.~\ref{HCOpMM}-\ref{HCNMM} and red and cyan lines in Fig.~\ref{PrjDep}).

With regards to the qualitative features in the zeroth moment maps, $\rm{HCN}$ and $\rm{HCO^+}$ exhibit a more complex behavior. For large values of the polar angle, both molecules are fair tracers of the column density, albeit both species peak off center in comparison to the column density peak. For such projection angles, $\rm{HCO^+}$ exhibits a more rapid decline than $\rm{HCN}$ in integrated intensity towards the outskirts of the cloud. For smaller values of the polar angle, both species trace the column density significantly better than $\rm{CO}$. However, $\rm{HCN}$ performs significantly better in the highest column density regions (column density contours > 50\%). In terms of mean values within the selected regions (red ellipses in Fig.~\ref{COMM}), $\rm{HCO^+}$ declines faster as a function of the projection angle than $\rm{HCN}$ (see red and cyan lines in the upper panel of Fig.~\ref{PrjDep}).

Among all species for which we have produced synthetic PPV data cubes, only $\rm{N_2H^+}$ is a good tracer of the $\rm{H_2}$ column density across all projection angles (left column in Fig.~\ref{N2HpMM}). We note, however, that this does not necessarily imply that $\rm{N_2H^+}$ also traces the $\rm{H_2}$ number-density peak. Specifically, due to the complex 3D structure of the cloud, the location of the column-density peak does not exactly coincide with the location of the number-density peak. Additionally, $\rm{N_2H^+}$ is the only species for which the integrated emission does not systematically decrease as a function of the projection angle (from face on to edge on). In contrast, by comparing the upper left and lower left panels of Fig.~\ref{N2HpMM}, we observe an increase in integrated emission by $\sim$20\% in the higher column density contours (> 70\%; dash-dotted blue lines), while the average value in the selected red regions shown in the right column of Fig.~\ref{COMM} remains roughly constant with projection angle (see magenta line in the upper panel of Fig.~\ref{PrjDep}).

A similar pattern to the zeroth moment maps is also seen in the second moment maps of our simulated PPV cubes (see right columns in Figs.~\ref{COMM},~\ref{HCOpMM}-\ref{N2HpMM} and the lower panel in Fig.~\ref{PrjDep}). That is, the dispersion of the LOS velocity is $\sim$20-30\% higher when the cloud is observed face on in comparison to when it is observed edge on. The same trend is evident in the synthetic PPV cubes of all simulated species. This is due to the fact that, even though the modeled cloud is supercritical, it still primarily collapses along magnetic field lines. As such, the velocity dispersion observed for small polar angles is primarily due to gravitational collapse. Interestingly enough however, these systematic motions along magnetic field lines are not particularly evident in the first moment maps of intermediate projection angles ($\phi = 22.5^\circ, 45^\circ$ and $67.5^\circ$ shown in the second, third and fourth rows of Fig.~\ref{COMM},~\ref{HCOpMM} - \ref{N2HpMM}, respectively), although a global gradient is observed. Instead, the first moment maps are primarily dominated by the turbulent motions within the cloud and exhibit both red and blue-shifted motions above and below the ``midplane'' of the cloud ($\eta = 0$ pc).

Additionally, the velocity dispersion derived from $\rm{HCN}$ synthetic observations is somewhat higher than that derived from $\rm{HCO^+}$ by $\sim0.03-0.05 \rm{km \ s^{-1}}$ for polar angles close to face on. We note that, since our simulation was performed assuming ideal MHD conditions, if $\rm{HCO^+}$ and $\rm{HCN}$ probed the same gas, the derived LOS velocity dispersion should be the same. This offset is also not constant with respect to the projection angle and when the cloud is seen edge on the derived velocity dispersion from $\rm{HCO^+}$ is slightly higher than $\rm{HCN}$. While such differences in the derived LOS velocity dispersion might seem insignificant, they are a significant fraction of the expected ion-neutral drift velocity for central number densities of $10^5~\rm{cm^{-3}}$. Given that these two species also probe different parts of the cloud (see discussion above) corroborate the findings by \cite{2023MNRAS.521.5087T} that these two species cannot be used to probe ambipolar diffusion in molecular clouds.

\subsection{Caveats}

A key assumption made throughout this study is that the gas is isothermal. Departures from isothermality can significantly influence the abundances of species in our chemical network, as most reaction rates are temperature dependent, and can also impact the observed line profiles. Line profiles will be affected by both temperature-driven changes in collisional excitation and through variations in the optical depth. For instance, an increase in temperature will increase the thermal linewidth, causing more cells to fall within one thermal linewidth away from each other (see Sec~\ref{RTSetup}), which will, in turn, lead to a higher cumulative optical depth along a given LOS.

Ammonia observations (e.g., \citeyearless{1996A&A...312..585L}; \citeyearless{2015ApJ...805..185S}; \citeyearless{2017ApJ...843...63F}) have established that, at the scales of cores, the temperature remains nearly constant, with only small variations on the order of 2-3 K. Additionally, theoretical work (e.g., \citeyearless{2001ApJ...557..736G}; \citeyearless{2002A&A...394..275G}) has demonstrated that variations in the abundances of the main coolants, such as $\rm{C^+}$ and $\rm{CO}$ (e.g., \citeyearless{1978ApJ...222..881G}), at high densities ($\gtrsim10^4~\rm{cm^{-3}}$) do not significantly affect the temperature. This is because gas-dust coupling ensures that any reduction in gas-phase cooling is balanced by energy exchange between gas and dust, and because molecular cooling transitions are often optically thick. Finally, chemo-dynamical simulations including heating and cooling processes (e.g., \citeyearless{2010MNRAS.404....2G}; \citeyearless{2012MNRAS.421....9G}) have shown that for visual extinctions $\gtrsim$4 mag, the gas is nearly isothermal. For reference, the maximum visual extinction in our simulation is $\sim$14 mag. At lower densities $\lesssim10^3~\rm{cm^{-3}}$, we should expect temperature variations of the factor of 2-3, as $\rm{CO}$ observations of molecular clouds also suggest (e.g., \citeyearless{2008ApJ...680..428G}).

Given the above discussion, we do not expect any realistic temperature variations to significantly impact our results on molecular abundances, particularly in the central regions of the cloud ($\gtrsim10^4~\rm{cm^{-3}}$), which are the focus of the present study. Furthermore, in the lower-density outskirts, the abundance of $\rm{CO}$ does not exhibit substantial variation except at the very edges of the cloud (see Fig.~\ref{SpatialChem}). This suggests that temperature variations in these regions should also remain limited, although this conclusion is constrained by the absence of explicit heating and cooling processes in our modeling. 

Regarding the line profiles, the flattened structure of our modeled cloud, implies that any temperature increases in the outskirts are likely to be more pronounced along the $z$ axis. Such variations would preferentially enhance the values in the zeroth and second moment maps when the cloud is viewed face on compared to edge on, due to the broader linewidths and increased excitation. In the edge-on case, temperature variations would primarily affect the gradients of the zeroth and second moment maps from the center to the outskirts\footnote{Even this effect should also be minimal as, by the time significant temperature variations occur at high $z$ values, the molecular abundance and $\rm{H_2}$ number density would have decreased sufficiently that the emission, in most molecules, would be insignificant.}. As such, the observed trends in the moment maps discussed here, should be robust and would likely be strengthened rather than diminished by any realistic temperature variations.


\section{Summary and conclusions}\label{discuss}

We have presented first results from a 3D ideal MHD turbulent chemo-dynamical simulation of a collapsing molecular cloud, where we simultaneously followed the evolution of 115 chemical species. By post-processing this simulation with a non-LTE radiative-transfer code we produced line-emission synthetic observations under various projection angles, including on and off-axis projections. Synthetic dust-polarization observations were also created for the same projection angles and are presented in a companion paper.

We find that the abundance distributions of various species in our chemo-dynamical simulation can qualitatively explain certain features observed in molecular-emission maps. Specifically, we find evidence for $\rm{CO}$ depletion in the densest regions of the cloud, that the $\rm{HCN}$ abundance remains high ($\sim 6.5\times10^{-10}$) even in lower-density regions ($\sim 10^3~\rm{cm^{-3}}$), and that $\rm{NH_3}$ is more peaked (in relation to the $\rm{H_2}$ number density) than $\rm{N_2H^+}$. However, quantitatively, there is a tendency for the abundances of most species in our simulation to be overpredicted in comparison to most observational estimates. We attribute this trend to the fact that the cloud collapses rapidly and, as a result, these species do not have enough time to deplete onto dust grains. Our results hint that the initial conditions used in this specific model might not be entirely representative of real molecular clouds and cores.

From our synthetic observations we found that the projection angle has a notable effect on the observed properties of the cloud. For three out of four simulated species ($\rm{CO}$, $\rm{HCO^+}$, and $\rm{HCN}$), there is a decrease in the values of their zeroth moment maps (by $\sim$30-50\%) as the polar angle increases. Only for $\rm{N_2H^+}$ the zeroth moment map remains roughly constant (with even a slight increase in the very central regions) from face on to edge on. For all species, the velocity dispersion measured, is higher when the cloud is observed face on rather than when it is seen edge on. This is to be expected, as the cloud is primarily collapsing along magnetic-field lines and we are therefore observing the effect of gravity. However, such motions are not easily distinguishable in the first moment maps of the simulated species as turbulence significant affects their overall structure.

In \greenhyperref{paperII}{Paper II} we show that column density maps exhibit an opposite trend compared to the zeroth and second moment maps with respect to the projection angle. That is, the column density is higher when the cloud is seen edge on compared to when it is seen face on. Such complications, related to the projection angle, can affect observational estimates of various quantities, including those for the magnetic field strength (calculated using the Davis-Chandrasekhar-Fermi method and its variants), since all three input quantities, $\delta v, \delta \theta$ and $\rm{\rho_{H_2}}$, are affected by the inclination angle. Additionally, the $\rm{H_2}$ column density and the zeroth moment maps of molecular emission are inversely related to the projection angle. That is, the values of $\rm{H_2}$ column density increase with an increasing polar angle whereas the values in the zeroth moment maps decrease. Therefore, the slope in the relation between integrated molecular emission and column density, as examined in recent observational studies \citep{2023A&A...679A.112T}, should also be affected by the projection angle.

\begin{acknowledgements}

We are grateful to K. Tassis for stimulating discussions. We are also grateful to the referee for their constructive feedback and suggestions which helped enhance the clarity and quality of our manuscript. A. Tritsis acknowledges support by the Ambizione grant no. PZ00P2\_202199 of the Swiss National Science Foundation (SNSF). S. Basu was supported by a Discovery grant from NSERC. C. Federrath acknowledges funding provided by the Australian Research Council (Discovery Projects DP230102280 and DP250101526). The software used in this work was in part developed by the DOE NNSA-ASC OASCR Flash Center at the University of Chicago. This research was enabled in part by support provided by SHARCNET (Shared Hierarchical Academic Research Computing Network) and Compute/Calcul Canada and the Digital Research Alliance of Canada. We further acknowledge high-performance computing resources provided by the Leibniz Rechenzentrum and the Gauss Centre for Supercomputing (grants~pr32lo, pr48pi and GCS Large-scale project~10391), the Australian National Computational Infrastructure (grant~ek9) and the Pawsey Supercomputing Centre (project~pawsey0810) in the framework of the National Computational Merit Allocation Scheme and the ANU Merit Allocation Scheme. We also acknowledge use of the following software: \textsc{Matplotlib} \citep{2007ComputSciEng.9.3}, \textsc{Numpy} \citep{2020Nat.585..357}, \textsc{Scipy} \citep{2020NatMe..17..261V} and the \textsc{yt} analysis toolkit \citep{2011ApJS..192....9T}.

\end{acknowledgements}

%
%

\appendix
\onecolumn

\section{Dense-gas tracers}\label{DenseTracers}

\begin{figure*}[h]
\hspace{4mm}
\includegraphics[width=0.965\columnwidth, clip]{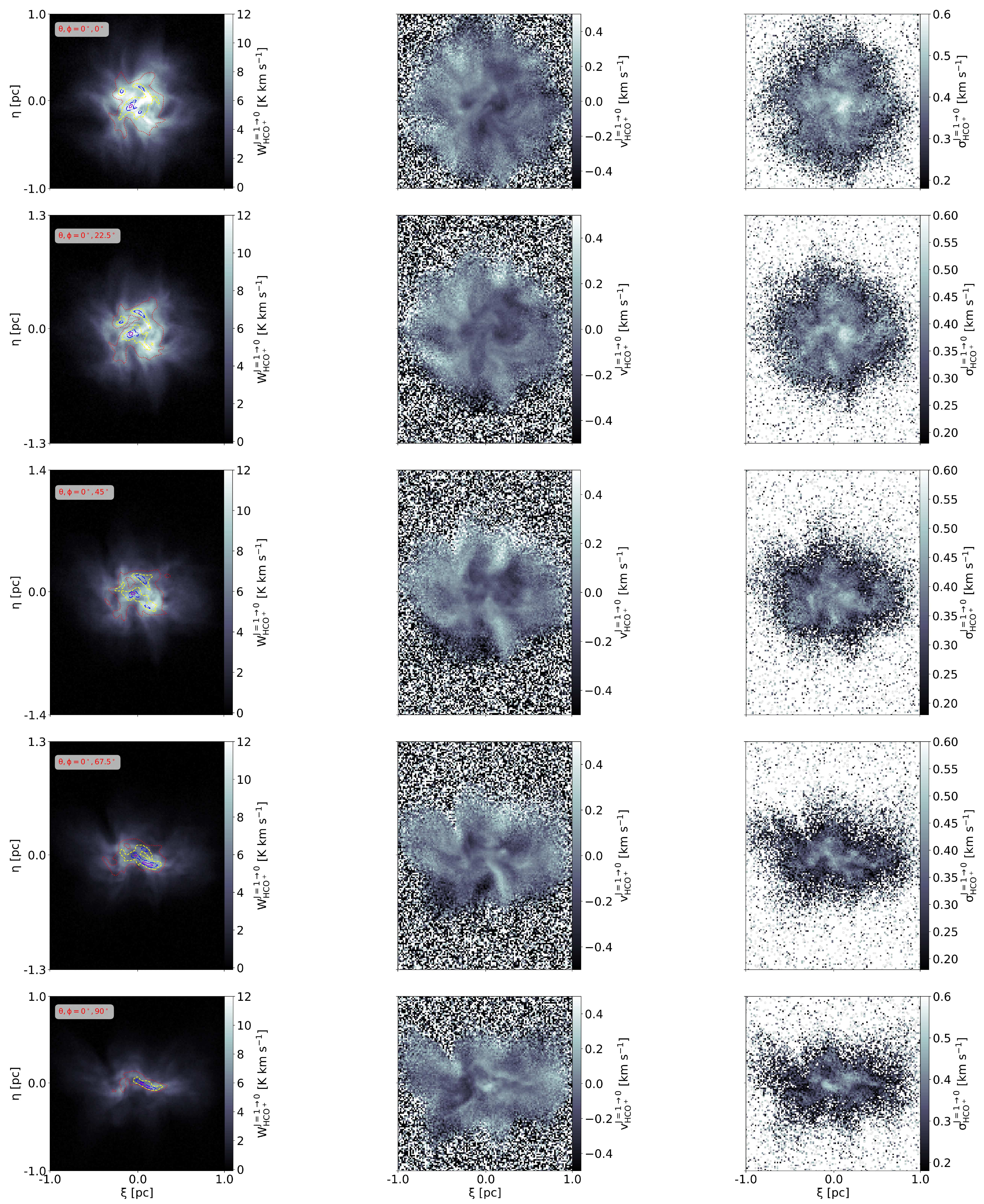}
\caption{Same as Fig.~\ref{COMM} but for the $\rm{HCO^+}~(J = 1\rightarrow0)$ transition.
\label{HCOpMM}}
\end{figure*}

\begin{figure*}
\hspace{4mm}
\includegraphics[width=0.965\columnwidth, clip]{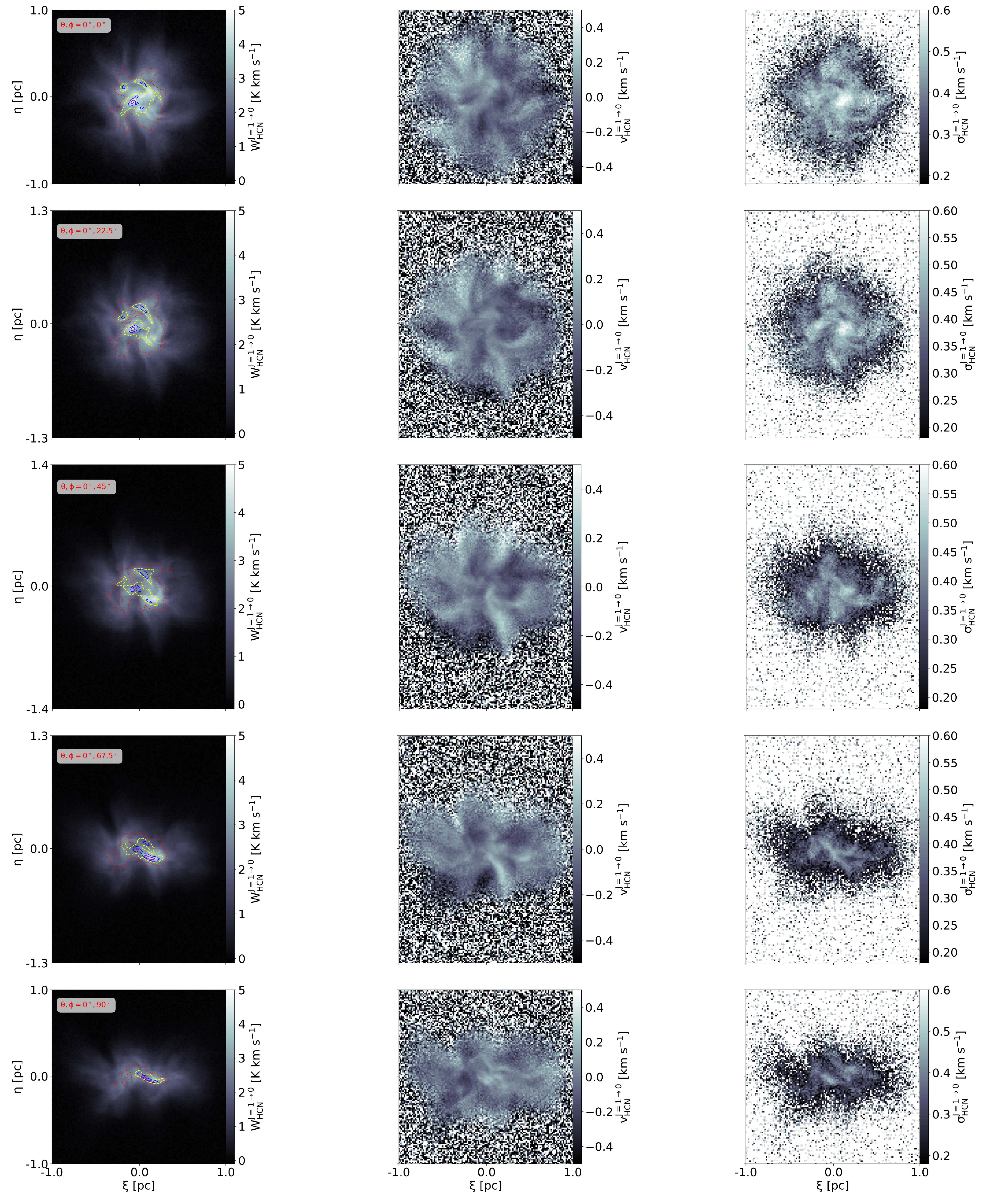}
\caption{Same as Fig.~\ref{COMM} but for the $\rm{HCN}~(J = 1\rightarrow0)$ transition.
\label{HCNMM}}
\end{figure*}

\begin{figure*}
\hspace{4mm}
\includegraphics[width=0.965\columnwidth, clip]{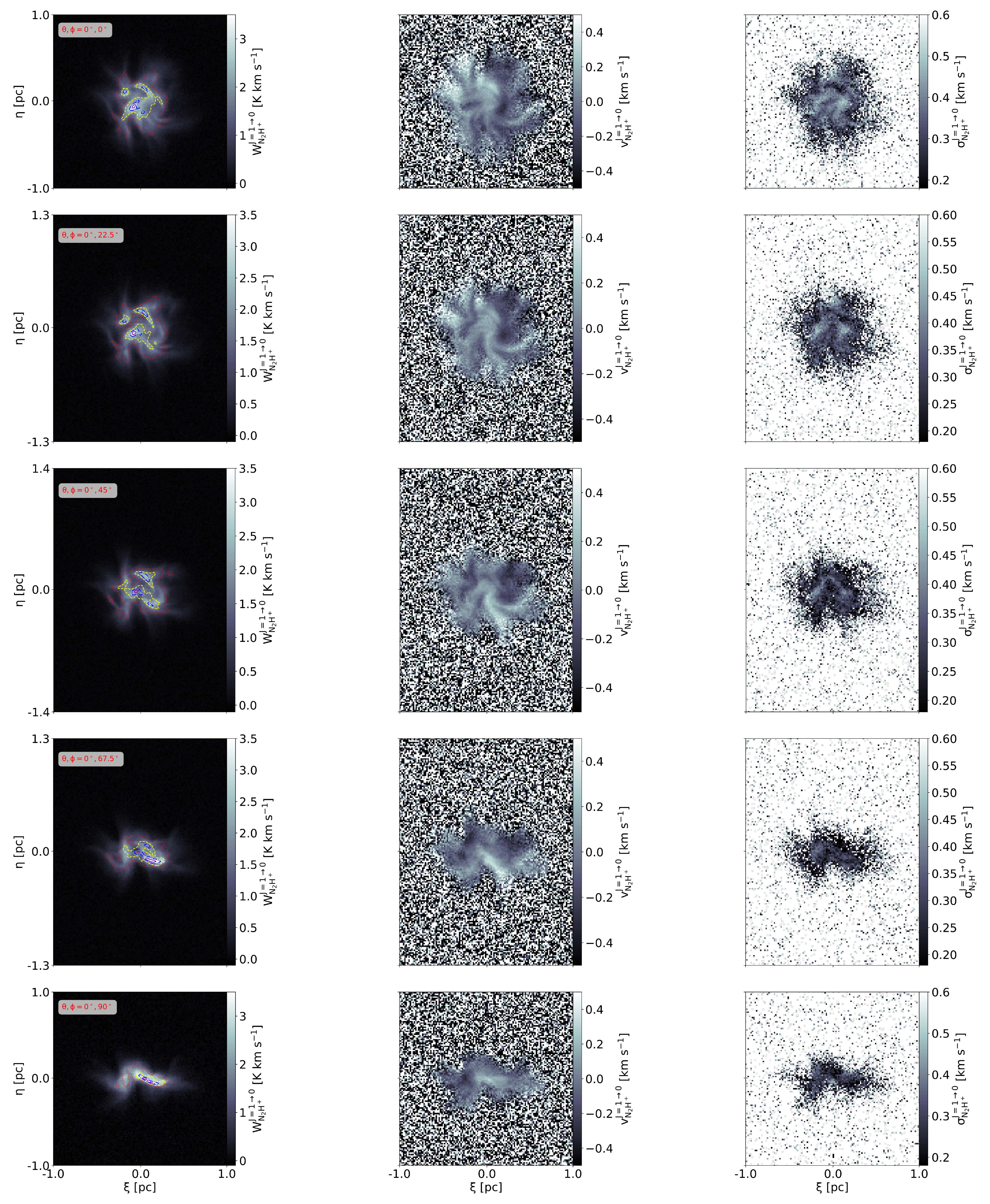}
\caption{Same as Fig.~\ref{COMM} but for the $\rm{N_2H^+}~(J = 1\rightarrow0)$ transition. In contrast to the other species studied here, $\rm{N_2H^+}$ is a good tracer of the column density for most projection angles.
\label{N2HpMM}}
\end{figure*}

In this Appendix we present moment maps for additional molecular species. Specifically, we have performed non-LTE radiative-transfer calculations for $\rm{HCO^+}$, $\rm{HCN}$ and $\rm{N_2H^+}$. In the interest of saving computational time, we have ignored the hyperfine structure for our $\rm{N_2H^+}$ radiative-transfer simulations. The corresponding moment maps are shown in Figs.~\ref{HCOpMM}, \ref{HCNMM}, \ref{N2HpMM}.

\end{document}